\documentclass[aps,preprint,showpacs,floats,epsf,epsfig,nofootinbib,12pt]{revtex4}
\usepackage{graphicx}
\usepackage{graphicx}
\usepackage{epsfig}
\usepackage{amsmath}
\usepackage{array}
\usepackage{epstopdf}
\usepackage{amstext}
\usepackage{amssymb}

\def\beq{\begin{eqnarray}}
\def\eeq{\end{eqnarray}}

\def\la{\langle}
\def\ra{\rangle}

\makeatletter

\newcommand{\Rmnum}[1]{\expandafter\@slowromancap\romannumeral #1@}
\makeatother

\begin{document}

\title{The decay of $\Lambda_b\rightarrow p~K^-$ in QCD factorization approach }

\vspace{1cm}

\author{Jie Zhu$^1$\footnote{zhujllwl@mail.nankai.edu.cn},
 Hong-Wei Ke$^{2}$\footnote{khw020056@hotmail.com}, and
 Zheng-Tao Wei$^1$\footnote{weizt@nankai.edu.cn} }

\affiliation{ $^{1}$ School of Physics, Nankai University, Tianjin 300071, China \\
  $^{2}$ School of Science, Tianjin University, Tianjin 300072, China }


\begin{abstract}

With only the tree level operator, the decay of $\Lambda_b\rightarrow pK$ is predicted
to be one order smaller than the experimental data. The QCD penguin effects should be
taken into account. In this paper, we explore the one-loop QCD corrections to the
decay of $\Lambda_b\to pK$ within the framework of QCD factorization approach. For the
baryon system, the diquark approximation is adopted. The transition hadronic matrix
elements between $\Lambda_b$ and $p$ are calculated in the light front quark model.
The branching ratio of $\Lambda_b\rightarrow pK$ is predicted to be about
$4.85\times 10^{-6}$ which is consistent with experimental data $(4.9\pm 0.9)\times 10^{-6}$.
The CP violation is about 5\% in theory.


\end{abstract}

\maketitle

\section{Introduction}

The weak decays of the heavy baryon $\Lambda_b$ provide an ideal place to extract information about the Cabibbo-Kobayashi-Maskawa (CKM) parameters and explore the mechanism of CP violation complementary to the B meson system. For the non-leptonic processes, the strong interaction dynamics is very complicated. Thus, these processes are also good probes to test different QCD models and factorization approaches. In early works of \cite{Wei:2009np,Ke:2007tg}, the weak decay of $\Lambda_b$ to $\Lambda_c$ and light baryons ($p$, $\Lambda$) are systematically studied. The hadronic transition matrix elements parameterized by form factors are calculated by use of a light-front quark model (LFQM) \cite{Terentev:1976jk, Jaus:1989au, Ji:1992yf, Cheng:1996if, Cheng:2003sm, Hwang:2006cua}. Since there are three valence quarks in a baryon, the quark-diquark picture was employed for simplification. It is found that the diquark approximation not only greatly simplifies the calculations, but also gives well theoretical predictions.

With a simple factorization hypothesis, many non-leptonic processes of $\Lambda_b$ to a light baryon and a meson are calculated in \cite{Wei:2009np}. The theory predictions of branching ratios are well consistent with the experiment data except one process of $\Lambda_b\rightarrow p~K^-$. The theory result is $Br(\Lambda_b\rightarrow p~K^-)=2.58\times 10^{-7}$, which is one order smaller than the data $(4.9\pm 0.9)\times 10^{-6}$ \cite{Agashe:2014kda}. What is the reason? In fact, the physics reason had been discussed in \cite{Wei:2009np}. The calculations are performed at the tree level. In most cases, the tree operator contribution is dominant. However, for the $\Lambda_b\rightarrow p~K^-$ process, the tree level contribution is suppressed by the CKM matrix elments $V_{ub}V^{*}_{us}$. For the penguin diagram, the main contribution comes from the loop where top quark is the dominant intermediate fermion. The CKM entry would be $V_{tb}V^{*}_{ts}$ which is almost 50 times larger than $V_{ub}V^{*}_{us}$. Thus even though there is a loop suppression of order $\alpha_s/4\pi$, it is compensated by the much larger CKM parameter, so the contributions from penguin diagrams are dominant. The effects of QCD penguin have been displayed in $B\to \pi K$ processes. For example, the process of $B^0\to K^+\pi^-$ is QCD penguin dominated and its branching ratio is $(1.94\pm 0.06)\times 10^{-5}$, while for a tree dominated process $B^0\to \pi^+\pi^-$ with $Br(B^0\to \pi^+\pi^-)=(5.15\pm 0.22)\times 10^{-6}$, the ratio is a factor of three smaller than that of $B^0\to K^+\pi^-$.

Using the method of perturbative QCD (pQCD) approach, $\Lambda_b\rightarrow p~K^{-}$ has been
calculated in \cite{Lu:2009cm}. The result is $1.82\times 10^{-6}$ in conventional pQCD approach and $2.02\times 10^{-6}$ in hybrid pQCD approach. We can see that it is smaller than a half of the experimental data $(4.9\pm 0.9)\times 10^{-6}$. In this paper, we will study the QCD corrections in the decay $\Lambda_b\rightarrow p~K^{-}$ at one-loop order within the framework of QCD factorization approach \cite{Beneke:1999br, Beneke:2000ry, Beneke:2001ev, Beneke:2003zv}. This factorization approach provides a systematic method to treat the non-factorizable QCD effects. It has been widely applied into many B meson non-leptonic processes. We will employ this approach into the heavy baryon decays, the $\Lambda_b\rightarrow p~K^{-}$ process in this study.

The paper is organized as follows: In Section II, we list the effective Hamiltonian for the transition  $\Lambda_b\rightarrow p~K^-$, give the QCD factorization approach to $\Lambda_b\rightarrow p~K^-$, the decay rate, and then discuss CP asymmetry
and the relation to decay of $\bar{B^0}\rightarrow K^- \pi^+$.  In Section III,  we will give the numerical calculations. In Section IV, a discussion and conclusion is provided.

\section{The decay $\Lambda_b\rightarrow p\, K^-$ }
\subsection{Effective Hamiltonian for  $\Lambda_b\rightarrow p\, K^-$}

In the decay $\Lambda_b\rightarrow p\, K^-$, the initial $\Lambda_b$ and final $p$ are baryons with three valence quarks. When the diquark picture is employed, i.e. the inner quark structure of $\Lambda_b$ is $b[ud]$ and $p$ is $u[ud]$ where $[ud]$ is a scalar diquark in this case and acted as a spectator.  The effective Hamiltonian $H_{eff}$ for $b\rightarrow s$ transitions can be written by:
 \begin{eqnarray}
  \mathcal{H}_{eff}=\frac{G_F}{\sqrt{2}}\sum\limits_{q=u,c}V_{qb}V^{*}_{qs}\left(C_1O^{q}_1+
  C_2O^{q}_2+\sum\limits^{10}_{i=3}C_{i}O_{i}+C_{7\gamma}O_{7\gamma}+C_{8g}O_{8g}\right),
 \end{eqnarray}
where $C_i$ are the Wilson coefficients evaluated at the renormalization scale $\mu$; the current-current operators $O_1^{u}$ and $O_2^{u}$ read
 \begin{eqnarray}
 O^{u}_1=\bar{s}_{\alpha}\gamma^{\mu}L u_{\alpha}\cdot\bar{u}_{\beta}\gamma_{\mu}L b_{\beta}, ~~~~~ &O^{u}_2=\bar{s}_{\alpha}\gamma^{\mu}L u_{\beta}\cdot\bar{u}_{\beta}\gamma_{\mu}L b_{\alpha},
 \end{eqnarray}
The usual tree-level W-exchange contribution in the effective
theory corresponds to $O_1$ and $O_2$ emerges due to the QCD
corrections. The QCD penguin operators $O_3-O_6$ are
 \begin{eqnarray} \nonumber
 O_3=\bar{s}_{\alpha}\gamma^{\mu}L b_{\alpha}\cdot\sum\nolimits_{q'}\bar{q}'_{\beta}\gamma_{\mu}Lq'_{\beta}, ~~~~~ & O_4=\bar{s}_{\alpha} \gamma^{\mu}L b_{\beta}\cdot\sum\nolimits_{q'}\bar{q}'_{\beta}\gamma_{\mu}L
  q'_{\alpha},\\[8pt]
 O_5=\bar{s}_{\alpha}\gamma^{\mu}L b_{\alpha}\cdot\sum\nolimits_{q'}\bar{q}'_{\beta}\gamma_{\mu}R
  q'_{\beta}, ~~~~~ &
 O_6=\bar{s}_{\alpha}\gamma^{\mu}L b_{\beta}\cdot\sum\nolimits_{q'}\bar{q}'_{\beta}\gamma_{\mu}R
  q'_{\alpha},
\end{eqnarray}
They contribute in order $\alpha_s$ through the initial values of the Wilson coefficients at $\mu\approx M_W$ \cite{Buras:1991jm} and operator mixing due to the QCD correction \cite{Shifman:1975tn}. Some operators $O_7,\ldots,O_{10}$ which arise from the electroweak-penguin diagrams are
 \begin{eqnarray}
 \nonumber O_7=\frac{3}{2} \bar{s}_{\alpha}\gamma^{\mu}L b_{\alpha}\cdot\sum\nolimits_{q'}e_{q'}\bar{q}'_{\beta}\gamma_{\mu}R q'_{\beta}, ~~~~~
 &&O_8=\frac{3}{2} \bar{s}_{\alpha}\gamma^{\mu}L b_{\beta}\cdot\sum\nolimits_{q'}e_{q'}\bar{q}'_{\beta}\gamma_{\mu}R q'_{\alpha},\\[8pt]
 O_9=\frac{3}{2} \bar{s}_{\alpha}\gamma^{\mu}L b_{\alpha}\cdot\sum\nolimits_{q'}e_{q'}\bar{q}'_{\beta}\gamma_{\mu}L q'_{\beta}, ~~~~~
 &&O_{10}=\frac{3}{2} \bar{s}_{\alpha}\gamma^{\mu}L b_{\beta}\cdot\sum\nolimits_{q'}e_{q'}\bar{q}'_{\beta}\gamma_{\mu}L q'_{\alpha},
 \end{eqnarray}
Here $\alpha$ and $\beta$ are the SU(3) color indices. There are still two operators
 \begin{eqnarray}
 O_{7\gamma}=\frac{-e}{8\pi^2}m_b\bar{s}\sigma_{\mu\nu}(1+\gamma_5)F^{\mu\nu}b, ~~~~~
 &&O_{8g}=\frac{-g_s}{8\pi^2}m_b\bar{s}\sigma^{\mu\nu}RG^{\mu\nu}b.
 \end{eqnarray}
$O_{7\gamma}$ and $O_{8g}$ are the electromagnetic, chromomagnetic dipole operators and $G^{\mu\nu}$ denotes the gluonic field strength tensor. In the above equations, $L$ and $R$ are the left- and right-handed projection
operators with $L=1-\gamma_5$ and $R=1+\gamma_5$  respectively. The sum over $q'$ runs over the quark fields that are active at the scale $\mu=O(m_b)$, i.e. $q'={u,d,s,c,b}$. The difficult problem is how to calculate the hadronic matrix elements of the local effective operators.

\subsection{$\Lambda_b\rightarrow p ~K^-$ in QCD factorization approach}

\begin{figure}[h]
\centering
\includegraphics[scale=0.9]{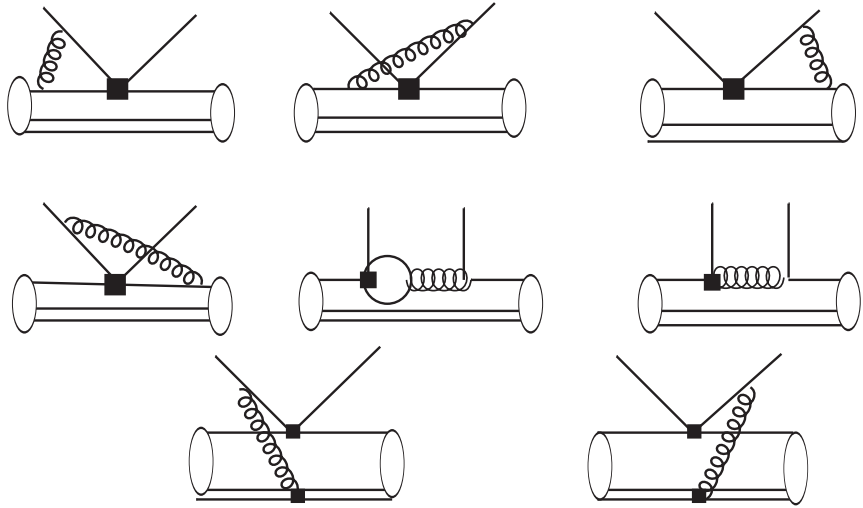}
\caption{Order $\alpha_s $ corrections to the hard-scattering kernels $T^{1}$ (first two
 rows)and $T^{2}$ (last row)}
\end{figure}

The naive factorization neglects the strong interactions between the final $K$ meson and two baryons. It is necessary to consider the non-factorizable contributions. There are several approaches which are beyond the naive factorization. In this study, we use the method called  QCD factorization approach \cite{Beneke:1999br, Beneke:2000ry, Beneke:2001ev, Beneke:2003zv}. The QCD factorization proves that in the heavy quark limit, the decay amplitude can be factorized into a product of hard scattering kennel and non-perturbative part. The $K$ meson and proton are both light hadron and energetic.  The interaction between them should be caused by large momentum transfer. Although the proof is given for the B meson case, it would be valid for the baryon system, too. Since we adopt the diquark approximation, the complications caused by more valence quarks nearly vanish. The diquark, as a whole, seems to be a light quark (it should be noted that the diquark in our case is a scalar while quark is a fermion). Thus, we assume that QCD factorization can be applicable to $\Lambda_b\rightarrow p\, K^-$.

The diagram for the $\alpha_s$ order QCD corrections to $\Lambda_b\rightarrow p\, K^-$ is plotted in Fig. 1. The fist two rows represents one-loop vertex corrections and $\alpha_s$ corrections to electromagnetic, chromomagnetic dipole operators. The last row represents the hard spectator scattering. At present, we don't know the wave function for a baryon with a quark and a diquark. One may use a meson like wave function, but a quantity like the decay constant is unknown. Thus, we will neglect the hard spectator contributions. After this simplification, the decay amplitude of $\Lambda_b\rightarrow p\, K^-$ can be written by
 \beq
 \langle p\, K^-|O_i|\Lambda_b\rangle=F^{\Lambda_b\rightarrow p}~T^{1}_{i}~*~f_K\Phi_K.
 \eeq
Here, $F^{\Lambda_b\rightarrow p}$ represents the $\Lambda_b\rightarrow p$ form factors which will be defined below; $*$ represents a convolution in the light-cone momentum fraction space; $T^{1}_{i}$ represents the four-quark hard scattering kernel; $\Phi_K$ represents the kaon meson light-cone wave function.

In QCD factorization, the amplitude $\Lambda_b\rightarrow p\, K^-$ is obtained as
 \begin{eqnarray}\label{eq:Mt}
 \mathcal{M}=&&\frac{G_F}{\sqrt{2}}\left\{V_{ub}V_{us}^{*}a_1+{V_{qb}V_{qs}^{*}}
 \left[a_4^q+a_{10}^q+R\left(a_6^q+a_8^q\right)\right]\right\}\nonumber\\[8pt]
 &&\times\langle p\mid\bar{u}\gamma_{\mu}L
 b\mid\Lambda_b\rangle\langle K^-\mid\bar{s}\gamma^{\mu}L u\mid 0\rangle,
 \end{eqnarray}
Here, a summation over $q= u,c$ is implicit. The $a_i$ are written as
 \begin{eqnarray}
 a_1&=&C_1+\frac{C_2}{N_c}\left[1+\frac{C_F\alpha_s}{4\pi}V_K\right],\nonumber\\[8pt]
 a_4^q&=&C_4+\frac{C_3}{N_c}\left[1+\frac{C_F\alpha_s}{4\pi}V_K\right]+\frac{C_F\alpha_s}{4\pi}
 \frac{P^q_{K,2}}{N_c},\nonumber\\[8pt]
 a_6^q&=&C_6+\frac{C_5}{N_c}\left(1-6\frac{C_F\alpha_s}{4\pi}\right)+\frac{C_F\alpha_s}{4\pi}
 \frac{P^q_{K,3}}{N_c},\nonumber\\[8pt]
 a_8^q&=&C_8+\frac{C_7}{N_c}\left(1-6\frac{C_F\alpha_s}{4\pi}\right)+\frac{\alpha}{9\pi}
 \frac{P^{q,EW}_{K,3}}{N_c},\nonumber\\[8pt]
 a_{10}^q&=&C_{10}+\frac{C_9}{N_c}\left[1+\frac{C_F\alpha_s}{4\pi}V_K\right]+\frac{\alpha}{9\pi}
 \frac{P^{q,EW}_{K,2}}{N_c}.
\end{eqnarray}
where $C_i\equiv C_i(\mu)$, $\alpha_s\equiv\alpha_s(\mu)$, $C_F=(N^2_c-1)/(2N_c)$, and $N_c=3$.
The quantities $V_K$,  $P^q_{K,2}$, $P^q_{K,3}$, $P^{q,EW}_{K,2}$, and $P^{q,EW}_{K,3}$ are
hadronic parameters that contain all nonperturbative dynamics. Their expressions are given
in \cite{Beneke:2001ev}. These quantities consist of convolutions of hard-scattering kernels with
meson distribution amplitudes. The term $V_K$ denotes the vertex corrections, $P^q_{K,2}$ and $P^q_{K,3}$ denote QCD penguin corrections and the contributions from the dipole operators. For penguin terms, the subscript 2 or 3 indicates the twist of the corresponding projections.

\subsection{The decay rate}

In Eq. (\ref{eq:Mt}), the first factor $\la p | J_\mu |\Lambda_b\ra$ in the hadronic matrix element is parameterized by form factors. The calculations of these non-perturbative form factors is one essential work of hadron physics. The form factors for the weak transition $\Lambda_b\rightarrow p$ are defined in the standard way as
 \begin{eqnarray}\label{s1}
 &&\la p(P') \mid \bar u\gamma_{\mu}
 (1-\gamma_{5})b \mid \Lambda_{b}(P) \ra  \nonumber \\
 &&= \bar{u}_{p}(P') \left[ \gamma_{\mu} f_{1}(q^{2})
 +i\sigma_{\mu \nu} \frac{ q^{\nu}}{M_{\Lambda_{b}}}f_{2}(q^{2})
 +\frac{q_{\mu}}{M_{\Lambda_{b}}} f_{3}(q^{2})
 \right] u_{\Lambda_{b}}(P) \nonumber \\
 &&-\bar u_{p}(P')\left[\gamma_{\mu} g_{1}(q^{2})
  +i\sigma_{\mu \nu} \frac{ q^{\nu}}{M_{\Lambda_{b}}}g_{2}(q^{2})+
  \frac{q_{\mu}}{M_{\Lambda_{b}}}g_{3}(q^{2})
 \right]\gamma_{5} u_{\Lambda_{b}}(P),
 \end{eqnarray}

The second factor of matrix element in Eq. (\ref{eq:Mt}) defines the decay constants
as follows
 \beq\label{p1}
 \la K^-(P)|A_{\mu}|0\ra&=&f_K P_{\mu}.
 \eeq
In the above definition, we omit a factor $(-i)$ for the  pseudoscalar meson decay constant for simplification.

Substituting the expressions of $\langle K^-\mid\bar{s}\gamma^{\mu}(1-\gamma^5)u\mid 0 \rangle$ and
$\langle p\mid\bar{u}\gamma_{\mu}(1-\gamma^5)b\mid\Lambda_b\rangle$ one can
obtain the decay amplitude of $\Lambda_b\rightarrow p~ K^-$ as
\begin{equation}
\mathcal{M}(\Lambda_b\rightarrow
p~K^-)=\bar{u}_p(A+B\gamma_5)u_{\Lambda_b},
\end{equation}
with
\begin{eqnarray}
A=\lambda f_K(M_{\Lambda_b}-M_p)f_1(M_K^2),&& B=\lambda
f_K(M_{\Lambda_b}+M_p)g_1(M_K^2),\nonumber
\end{eqnarray}
where \begin{equation}
\lambda=\frac{G_F}{\sqrt{2}}\left\{V_{ub}V_{us}^{*}a_1+{V_{qb}V_{qs}^{*}}
 \left[a_4^q+a_{10}^q+R\left(a_6^q+a_8^q\right)\right]\right\}.\nonumber
\end{equation}
Then we get the decay rate of $\Lambda_b\rightarrow p~K^-$
\begin{equation}
\Gamma=\frac{p_c}{8\pi}\left[\frac{(M_{\Lambda_b}+M_p)^2-M_K^2}{M^2_{\Lambda_b}}\mid A\mid^2+\frac{(M_{\Lambda_b}-M_p)^2-M_K^2}{M^2_{\Lambda_b}}\mid B\mid^2\right].
\end{equation}
where $p_c$ is the proton momentum in the rest frame of
$\Lambda_b$.

\subsection{CP asymmetry and relation to decay of $\bar{B^0}\rightarrow \pi^+ K^- $}

The CP violation is defined in the same way as PDG book \cite{Agashe:2014kda} by
 \beq
 A_{CP}\equiv\frac{Br(\Lambda_b^0\rightarrow p~ K^-)-Br(\bar{\Lambda_b^0}\rightarrow \bar p~ K^+)}
 {Br(\Lambda_b^0\rightarrow p~ K^-)+Br(\bar{\Lambda_b^0}\rightarrow \bar p~ K^+)},
 \eeq
At the quark level, the CP violation is represented by $b$ quark decay minus $\bar b$ quark. The similar definition of CP violation for meson is
 \beq
 A_{CP}\equiv\frac{Br(\bar B^0\rightarrow f)-Br(B^0\rightarrow \bar f)}
 {Br(\bar B^0\rightarrow f)+Br(B^0\rightarrow \bar f)}.
 \eeq

Under the diquark approximation, the baryon is similar to a meson. In fact, at the quark level, $\Lambda_b\rightarrow p K^-$ has the same sub-processes $b\rightarrow su\bar{u}$ as that in $\bar{B^0}\rightarrow \pi^+ K^-$.  The amplitude of $\bar{B^0}\rightarrow \pi^+ K^-$ is
 \begin{eqnarray} \label{bkpi}
 M(\bar{B}^0\rightarrow \pi^+ K^-) = && \frac{G_F}{\sqrt{2}}\left\{V_{ub}V_{us}^{*}a_1+{V_{qb}V_{qs}^{*}}
 \left[a_4^q+a_{10}^q+R\left(a_6^q+a_8^q\right)\right]\right\}\nonumber\\[8pt]
 &&\times\langle \pi^+\mid\bar{u}\gamma_{\mu}L b\mid\bar{B}^0\rangle\langle K^-\mid\bar{s}\gamma^{\mu}L u\mid
 0\rangle.
 \end{eqnarray}
Compare it to Eq. (\ref{eq:Mt}), we can obtain a relation between the baryon and meson processes,
 \begin{eqnarray}\label{eqcd}
 Br(\Lambda_b\rightarrow p~K^- ) =&&Br^{\rm Exp}(\bar{B}^0\rightarrow
 \pi^+ K^- )\times\frac{Br^{\rm tree}(\Lambda_b\rightarrow p~
 K^-)}{Br^{\rm tree}(\bar{B}^0\rightarrow \pi^+ K^- )} .
 \end{eqnarray}
Where the "tree" represents the branching ratio with only the tree operator contribution.
This relation will be used to estimate the branching ratio of $\Lambda_b\rightarrow p~K^-$ from the
meson process $\bar{B}^0\rightarrow K^- \pi^+$. About the CP violation, under the above assumption, $A_{CP}$ in the two processes should be equal.

\section{Numerical Results}

At first, we list some parameters used in the numerical calculations. The input parameters are taken from \cite{Agashe:2014kda} and the previous works.
 \begin{eqnarray}
 \begin{array}{l l l}
  m_u=0.3 ~\textrm{GeV}, &\qquad m_s=0.45 ~\textrm{GeV}, &\qquad m_c =1.3 ~\textrm{GeV},\nonumber\\
  m_K=0.4937 ~\textrm{GeV}, &\qquad m_b=4.4 ~\textrm{GeV},  &\qquad m_{[ud]} =0.5~ \textrm{GeV},\nonumber\\
  M_{\Lambda_b}=5.619 ~\textrm{GeV}, &\qquad M_p=0.938 ~\textrm{GeV},&\qquad m_B =5.280~\textrm{GeV},\nonumber\\
  m_{\pi}=0.1396~\textrm{GeV}, &\qquad f_K =0.160~\textrm{GeV}&\qquad F_0^{B\rightarrow\pi}(0) =0.3~.
 \end{array}
 \end{eqnarray}
The above quark masses of $u,~d$ are the constitute masses which are used in the LFQM. While for the current quark masses, $m_u=2.3~\textrm{MeV}$ and $m_s=95~\textrm{MeV}$.

Following \cite{Wei:2009np}, we recalculate the from factors of $\Lambda_b\rightarrow p$ in the LFQM. The form factors at different $q^2$ are parametrized in a three-parameter form as
\begin{equation}
F(q^2)=\frac{F(0)}{\left(1-\frac{q^2}{M_{\Lambda_b}^2}\right)[1-a\left(\frac{q^2}
 {M_{\Lambda_b}^2}\right)+b\left(\frac{q^2}{M_{\Lambda_b}^2}\right)^2]}~.
\end{equation}
where the fitted values of $a$, $b$, and $F(0)$ are given in Table I.
\begin{table}[h]
\caption{The value of $a$, $b$ and $F(0)$.}
\begin{center}
\begin{tabular}{c c c  c}\hline\hline
 $~~~~~~~~~~F~~~~~~~~~~$   & ~~~~~~~~~~$F(0)$~~~~~~~~~~
  & ~~~~~~~~~~$a$~~~~~~~~~~  & ~~~~~~~~~~$b$~~~~~~~~~~  \\\hline
 $f_1$ & 0.1131  & 1.70 & 1.60 \\\hline
 $f_2$ & -0.0356 & 2.50 & 2.57 \\\hline
 $g_1$ & 0.1112  & 1.65 & 1.60 \\\hline
 $g_2$ & -0.0097 & 2.80 & 2.70 \\\hline\hline
\end{tabular}\\[10pt]
\end{center}
\end{table}
Our results reproduce those given in \cite{Wei:2009np}.

For Wilson coefficients $C_i$, we use the leading order (LO) results as given in \cite{Beneke:2001ev} and list them in Table II. As for the CKM matrix elements, we adopt the Wolfenstein parametrization beyond the LO from \cite{Buras:1998raa}:
 \begin{eqnarray}
 V_{ud}=&&1-\frac{1}{2}\lambda^2-\frac{1}{8}\lambda_4+\mathcal {O}(\lambda^6),~~~~~
  V_{us}=\lambda+\mathcal {O}(\lambda^7),~~~~~ V_{ub}=A\lambda(\rho-i\eta),\nonumber\\
 V_{cd}=&&-\lambda+\frac{1}{2}A^2\lambda^5[1-2(\rho+i\eta)]+\mathcal {O}(\lambda^7),\nonumber\\
 V_{cs}=&&1-\frac{1}{2}\lambda^2-\frac{1}{8}\lambda^4(1+4A^2)+\mathcal {O}(\lambda^6),~~~~~
  V_{cb}=A\lambda^2+\mathcal {O}(\lambda^8),\nonumber\\
 V_{td}=&&A\lambda^3[1-(\rho+i\eta)(1-\frac{1}{2}\lambda^2)]+\mathcal {O}(\lambda^7),\nonumber\\
 V_{ts}=&&-A\lambda^2+\frac{1}{2}A(1-2\rho)\lambda^4-i\eta A\lambda^4+\mathcal {O}(\lambda^6),\nonumber\\
 V_{tb}=&&1-\frac{1}{2}A^2\lambda^4+\mathcal {O}(\lambda^6).
\end{eqnarray}
Here, we take the value $A=0.822$, $\lambda=0.22535$, $\rho=0.155$,
$\eta=0.358$.

By use of the above input parameters, we can get the Wilson coefficients $a_i$ which is relevant to the process of $\Lambda_b\rightarrow p~K^-$ with the $\alpha_s$ order QCD corrections. The numerical results are given in Table III. Considering the theoretical uncertainties, our results are consistent with those in \cite{Beneke:2001ev}. The small difference can be ascribed to the input parameters and the hard spectator contributions we neglected. Although the scale $\mu$ dependence of the Wilson coefficients $a_i$ is reduced compared to the LO ones, there is still effect which is not negligible. This dependence implies the importance of higher order effects.

\begin{table}
\caption{The Wilson coefficients $C_i$ in LO.}
\begin{center}
\begin{tabular}{c c c c c c c}\hline\hline
&$C_1$&$C_2$&$C_3$&$C_4$&$C_5$&$C_6$\\\hline
$\mu=m_b/2$~~&~~~1.185~~~&~~~-0.387~~~&~~~0.018~~~&~~~-0.038~~~&~~~0.010~~~&~~~-0.053~~~\\\hline
$\mu=m_b$&1.117&-0.268&0.012&-0.027&0.008&-0.034\\\hline
$\mu=2m_b$&1.074&-0.181&0.008&-0.019&0.006&-0.022\\\hline\hline
&$C_7/\alpha$&$C_8/\alpha$&$C_9/\alpha$&$C_{10}/\alpha$&&\\\hline
$\mu=m_b/2$&-0.012&0.045&-1.358&0.418&&\\\hline
$\mu=m_b$&-0.001&0.029&-1.276&0.288&&\\\hline
$\mu=2m_b$&0.018&0.019&-1.212&0.193&&\\\hline
\end{tabular}
\end{center}
\end{table}

\begin{table}
 \caption{The numerical values of $a_i$ in QCD factorization. }
 \begin{center}
 \begin{tabular}{c|c|c|c}\hline\hline
 &$\mu=m_b/2$&$\mu=m_b$&$\mu=2m_b$\\\hline
 ~~~~~$a_1~~~~~$   & $ 1.089+0.047i$ & $ 1.064+0.026i$ & $1.044+0.015i$ \\
 $a^u_4$ & $-0.033-0.018i$ & $-0.031-0.016i$ & $-0.029-0.014i$ \\
 $a^c_4$ & $-0.034-0.006i$ & $-0.036-0.005i$ & $-0.033-0.005i$ \\
 $a^u_6$ & $-0.049-0.018i$ & $-0.038-0.015i$ & $-0.031-0.013i$ \\
 $a^c_6$ & $-0.054-0.007i$ & $-0.041-0.006i$ & $-0.034-0.006i$ \\
 $a^u_8$ & $ 3.3\times10^{-4}$ & $(1.9-0.6i)\times10^{-4}$ & $ (0.9-1.0i)\times10^{-4}$ \\
 $a^c_8$ & $ 3.2\times10^{-4}$ & $(1.8-0.3i)\times10^{-4}$ & $ (0.7-0.5i)\times10^{-4}$ \\
 $a^u_{10}$& ~~$(6.4+12.9i)\times10^{-4}$~~ &~~ $(2.3+9.1i)\times10^{-4}$~~ & ~~$(-1.8+6.6i)\times10^{-4}$~~\\
 $a^c_{10}$& $(6.4+13.0i)\times10^{-4}$ & $(2.2+9.4i)\times10^{-4}$&$(-2.0+7.2i)\times10^{-4}$\\\hline
 \end{tabular}
 \end{center}
 \end{table}

\begin{table}
\caption{The branching ratios of $\Lambda_b\rightarrow p~K^-$.}
\begin{center}
\begin{tabular}{c |c |c | c}\hline\hline
 & ~~~~~$\mu=m_b/2$~~~~~ & ~~~~~$\mu=m_b$~~~~~   & ~~~~~$\mu=2m_b$~~~~~ \\ \hline
 ~~~QCD factorization~~~ & $ 4.85\times10^{-6} $ & $3.35\times 10^{-6}$ & $2.57\times10^{-6}$\\\hline
 Estimation from meson data & \multicolumn{3}{c}{$ 4.82\times10^{-6}$}\\\hline
 Experimental data & \multicolumn{3}{c}{$(4.9\pm 0.9)\times10^{-6}$}\\ \hline\hline
\end{tabular}
\end{center}
\end{table}

Now, we can obtain the branching ratio of $\Lambda_b\rightarrow p~K^-$. The predicted results at different scale $\mu$ are listed in Table IV. As discussed above, the results have an un-negligible dependence on the choice of scale $\mu$. The higher the scale is, the lower the prediction is. The  result at $\mu= m_b/2$ give prediction of $4.85\times10^{-6}$ which is very well with the recent LHCb data $(4.9\pm 0.9)\times10^{-6}$. The good coincidence indicates that $\mu=m_b/2$ is more appropriate. From phenomenological point of view, $m_b$ is the largest scale in $b$ quark decay subprocess and each quark in the final hadrons does not carry the total momentum.  The momentum transfer between dirrerent quarks should be smaller than $m_b$. So the choice of $\mu$ at $\mu=m_b/2$ is more reasonable than at $m_b$.

Under the assumption by neglecting the strong interactions with the spectator quark (diquark for the baryon), $\Lambda_b\rightarrow p~K^-$ contains the same strong dynamics with $\bar{B}^0\rightarrow \pi^+ K^-$. We can use the data of meson process to extract the strong interaction information.
The advantage of this method is that the scale $\mu$ dependence is eliminated and the theory uncertainties of QCD factorization approach is reduced by experiment. By the aid of of the experimental data of $Br(\bar{B}^0\rightarrow \pi^+ K^-)$ and the Eq. (\ref{eqcd}), we estimate the decay rate with $Br(\Lambda_b^0\rightarrow p~K^-)=4.82\times10^{-6}$. It coincides with the experimental measurement very well.

The results of CP violation is displayed in Table V. Contrary to the branching ratio, the numerical results of CP violation of $\Lambda_b \rightarrow p~K^-$ becomes smaller as the scale $\mu$ decreases. At scale $\mu=m_b/2$, the CP violation is about 5\%. The experimental data from LHCb is
$0.37\pm0.17\pm0.03$. The central value is several times larger than theory prediction.
Because the experimental error is still large, it's too early to give a conclusion whether the theory coincides with the experiment or not. It is interesting and necesaary to compare the CP violation to the meson case. The data of CP violation in $\bar{B^0} \rightarrow \pi^+ K^-$ is also provided in Table V for comparison. The value is $-0.080\pm0.007\pm0.003$ with a negative sign. In our calculations under the diquark approximation, the CP violation of $\bar{B}^0 \rightarrow \pi^+ K^- $ and $\Lambda_b \rightarrow p~ K^-$ should be equal. However, we see
that the experimental data for the two processes are quite different, especially the sign is opposite. In fact, the CP violation for the process of $\bar{B}^0 \rightarrow \pi^+ K^- $  in the QCD factorization approach is a challenging problem for a long time. The theory prediction is not only inconsistent with the experiment data but also is wrong in sign.

\begin{table}
\caption{The CP violation $A_{CP}(\Lambda_b \rightarrow p~K^-$).}
\begin{center}
\begin{tabular}{c |c |c | c}\hline\hline
 & ~~~~~$\mu=m_b/2$~~~~~ & ~~~~~$\mu=m_b$~~~~~   & ~~~~~$\mu=2m_b$~~~~~ \\ \hline
 QCD factorization & 0.049 & 0.076 & 0.095 \\ \hline
 Experimental data & \multicolumn{3}{c}{$0.37\pm0.17\pm0.03$}\\ \hline \hline
 $A_{CP}(\bar{B^0} \rightarrow \pi^+K^-)$&\multicolumn{3}{c}{$-0.080\pm0.007\pm0.003$}\\ \hline\hline
\end{tabular}
\end{center}
\end{table}

\section{Discussion and conclusion}

The weak decay of $\Lambda_b$ contains fruitful information of strong interaction and provides an
important probe to test different theory approaches. In this work, we extend the QCD factorization approach to the heavy baryon decays, in particular the process of $\Lambda_b^0\rightarrow p~K^-$.
The previous literature considers only the tree diagram contribution and the theory result is one order smaller than the experiment. The $\Lambda_b^0\rightarrow p~K^-$ is a type of $b\to s$ transition which the QCD penguin diagram contribution is more important than the tree diagram because of the CKM parameter enhancement. The QCD correction is calculated to $\alpha_s$ order and the Wilson coefficients at different renormalization scale are given. For the baryon, the diquark approximation is applied. The $\Lambda_b\to p$ form factors are calculated in the light-front quark model. The branching ratio of $\Lambda_b^0\rightarrow p~K^-$ is predicted to be $4.85\times10^{-6}$ at scale $\mu=m_b/2$. The theory coincides with the experimental data $(4.9\pm 0.9)\times10^{-6}$ very well.

From the coincidence of theory and experiment, we can obtain some conclusions as following. (1) The perturbative contribution is dominant. The success provides a confidence of applicability of QCD factorization method to the more complicated heavy baryon processes. (2) The choice of $\mu=m_b/2$ is appropriate. Because the largest scale is $m_b$ in the $b$ quark decay subprocess, the real momentum transfer cannot reach $m_b$ and should be smaller than it. (3) The diquark ansatz works very well. The diquark approximation not only lead to a clear physics picture but also a great simplification for the numerical calculations. From this study and the previous literatures on heavy baryon decays, we may say that the diquark is really a working ansatz.

The main theory uncertainties come from several origins: the choice of scale $\mu$,  the $\Lambda\to p$ form factors, the neglected hard spectator interaction and the non-perturbative power corrections. The problem of scale $\mu$ has been discussed in the article. Its scale dependence is not negligible. The higher loop corrections may help to reduce the dependence but are usually difficult to be realized. Although the $\Lambda\to p$ form factors depend on model calculations, the reliability can be fitted by experiemnt. In \cite{Wei:2009np}, it shows that our calculated $\Lambda\to p$ form factors give a well prediction for $\Lambda_b\to p~ \pi^-$: the theory result of the branching ratio is $3.15\times 10^{-6}$ and the experimental data is $(3.5\pm 0.6({\rm stat})\pm 0.9({\rm syst}))\times 10^{-6}$. Thus, the model-dependent form factors don't cause large theoretical uncertainties.

For the meson case, the hard spectator scattering contributes a leading power correction. It modifies the Wilson coefficient $a_5$ largely. For the coefficients $a_4$ and $a_6$ which is relevant to this study, the hard spectator correction is either numerically small (about 10\%) or absent. Thus, for the baryon case, the contribution from the hard spectator interaction is small. 
The weak annihilation contribution is power suppressed. At the realistic $m_b$ scale, it is necessary to consider its effect. The estimation of it suffers from the problem of end-point singularity. According to analysis in \cite{Beneke:2001ev}, the numerical values of annihilation correction is less than 25\% compared to the leading power term. Even this, the correction has included the chiral enhanced twist-3 contribution. For the baryon case, we might expect a similar small or even smaller weak annihilation contribution because no such chiral enhancement exist for the baryon of proton.

The higher power correction is usually difficult to calculate. The consistence of theory at $\mu=m_b/2$ with data indicates that the non-perturbative power correction is less important and the perturbative contribution is dominant. One can use the data from $\bar{B^0}\to \pi^+K^-$ to reduce the theory uncertainties in QCD factorization approach. By this way, we obtain the decay rate with $Br(\Lambda_b^0\rightarrow p~K^-)=4.82\times10^{-6}$ which coincides the experiment very well.

About the CP violation, it provides us a very different physics picture. Under the diquark approximation and neglecting the spectator interactions, the theory predicts CP violation at level of about 5\% for both the baryon process $\Lambda_b^0\rightarrow p~K^-$ and meson case $\bar{B^0}\to \pi^+K^-$. The origin of the strong phase in QCD factorization approach comes from the quark loop in the vertex corrections. For the meson case, the theory result is positive. But the experiment data is negative, about $-10$\%. This obvious inconsistence implies the importance of non-perturbative corrections for CP violation. For the baryon $\Lambda_b^0\rightarrow p~K^-$, the experiment data gives a very large result: $0.37\pm0.17\pm0.03$. In QCD factorization approach, the perturbative contribution cannot reach 10\%. Because it's quite difficult to estimate the non-perturbative corrections, the prediction of CP violation in theory is a challenging research. We hope the future LHCb data can provide us a more precise measurement of CP violation in $\Lambda_b^0\rightarrow p~K^-$ to improve the development of theory.

\section*{Acknowledgement}

This work is supported by the National Natural Science Foundation
of China (NNSFC) under the contract Nos. 11175091, 11375128.

\vspace{0.8cm}

{\it  Note added} ~~~~After we put the manuscript of this work on arXiv:1603.02800 [hep-ph], we
are noticed a similar research \cite{Hsiao:2014mua}. The authors had studied $\Lambda_b\to p~M$ within the generalized factorization approach. For the process of $\Lambda_b\to p~K$, the predicted branching ratio and direct CP violation are similar to us. But it should be noted that the form factors and the Wilson coefficients in the two works are different.

\appendix

\end{document}